\documentclass[aps,prl,reprint,amssymb,showpacs,onecolumn,superscriptaddress,notitlepage]{revtex4-1}
\usepackage{bm}
\usepackage{graphicx}
\usepackage{dcolumn}
\usepackage{braket}
\usepackage{natbib}
\usepackage{amsmath}
\usepackage{color}
\usepackage{ulem}
\usepackage{siunitx}
\usepackage[resetlabels]{multibib}

\definecolor{gray}{rgb}{0.7,0.7,0.7}
\definecolor{orange}{rgb}{1, 0.4, 0}
\definecolor{dgreen}{rgb}{0.0, 0.4, 0.0}
\definecolor{yblue}{rgb}{0.06, 0.3, 0.57}

\newcommand\ldsout{\bgroup\markoverwith{\textcolor{blue}{\rule[0.5ex]{2pt}{0.4pt}}}\ULon}

\begin{document}
\title{Electrically controlled waveguide polariton laser}
\author{D. G. Su\'arez-Forero}
\affiliation{CNR NANOTEC, Institute of Nanotechnology, Via Monteroni, 73100 Lecce, Italy}
\author{F. Riminucci}
\affiliation{Dipartimento di Matematica e Fisica, "Ennio de Giorgi", Universit\`{a} del Salento, Campus Ecotekne, via Monteroni, Lecce, 73100, Italy.}
\affiliation{Molecular Foundry, Lawrence Berkeley National Laboratory, One Cyclotron Road, Berkeley, California, 94720, USA}
\author{V. Ardizzone}
\email{v.ardizzone85@gmail.com}
\affiliation{CNR NANOTEC, Institute of Nanotechnology, Via Monteroni, 73100 Lecce, Italy}
\author{M. de Giorgi}
\affiliation{CNR NANOTEC, Institute of Nanotechnology, Via Monteroni, 73100 Lecce, Italy}
\author{L. Dominici}
\affiliation{CNR NANOTEC, Institute of Nanotechnology, Via Monteroni, 73100 Lecce, Italy}
\author{F. Todisco}
\affiliation{CNR NANOTEC, Institute of Nanotechnology, Via Monteroni, 73100 Lecce, Italy}
\author{G. Lerario}
\affiliation{CNR NANOTEC, Institute of Nanotechnology, Via Monteroni, 73100 Lecce, Italy}
\author{L. N. Pfeiffer}
\affiliation{PRISM, Princeton Institute for the Science and Technology of Materials, Princeton Unviversity, Princeton, NJ 08540}
\author{G. Gigli}
\affiliation{Dipartimento di Matematica e Fisica, "Ennio de Giorgi", Universit\`{a} del Salento, Campus Ecotekne, via Monteroni, Lecce, 73100, Italy.}
\author{D. Ballarini}
\affiliation{CNR NANOTEC, Institute of Nanotechnology, Via Monteroni, 73100 Lecce, Italy}
\author{D. Sanvitto}
\email{daniele.sanvitto@nanotec.cnr.it}
\affiliation{CNR NANOTEC, Institute of Nanotechnology, Via Monteroni, 73100 Lecce, Italy}

\begin{abstract}
Exciton-polaritons are mixed light-matter particles offering a versatile solid state platform to study many-body physical effects. In this work we demonstrate an electrically controlled polariton laser, in a compact, easy-to-fabricate and integrable configuration, based on a semiconductor waveguide. Interestingly, we show that polariton lasing can be achieved in a system without a global minimum in the polariton energy-momentum dispersion. The surface cavity modes for the laser emission are obtained by adding couples of specifically designed diffraction gratings on top of the planar waveguide, forming an in-plane Fabry-Perot cavity. It is thanks to the waveguide geometry, that we can apply a transverse electric field in order to finely tune the laser energy and quality factor of the cavity modes. Remarkably, we exploit the system sensitivity to the applied electric field to achieve an electrically controlled population of coherent polaritons. The precise control that can be reached with the manipulation of the grating properties and of the electric field provides strong advantages to this device in terms of miniaturization and integrability, two main features for the future development of coherent sources from polaritonic technologies.
\end{abstract}

\maketitle
\section{Introduction}
A semiconductor system in which a photon emitted from an active medium has a larger probability of being reabsorbed than that of escaping out of the optical resonator is called to be in strong coupling. This condition accounts for the formation of the so called exciton polariton: a light-matter quasi-particle resulting from the hybridization between an electromagnetic cavity mode and an exciton dipole in a semiconductor \cite{Kavokin2008}. Since the first observation \cite{PhysRevLett.69.3314}, polaritonic systems have become a suitable platform to study fundamental physical phenomena; effects such as optical parametric oscillations \cite{Baumberg2000}, bistability \cite{Baas2004}, Bose-Einstein condensation \cite{Kasprzak2006,Balili2007}, superfluidity \cite{Amo2009,Lerario2017} and quantum vorticity \cite{Lagoudakis2008} are some of the most intriguing phenomena that have been demonstrated. Peculiar features of polaritons such as long coherence time and high nonlinearities are chased for the realization of integrated optical elements \cite{Sanvitto2016}. Experimental devices such as polariton transistors \cite{Ballarini2013} and routers \cite{Marsault2015} have indeed shown their viability for all-optical logic systems.

The so called ``polariton condensate'' \cite{Imamoglu1996}, is one of the most paradigmatic effects observed in strongly coupled systems. This phenomenon is observed when a phase transition to a coherent state of polaritons takes place above a critical density, without any population inversion. It has been demonstrated in systems such as semiconductor, organic or hybrid microcavities \cite{Bajoni2008,Chang2019,Christopoulos2007,Christmann2008,Kena-Cohen2010,Daskalakis2014,Guillet2011,Lu2012} and photonic crystal cavities \cite{Seo2007,Azzini2011,Nomura2009}. On the contrary, if the system undergo the transition to the weak coupling regime before the condensate is formed, upon increasing power, a standard lasing effect starts to kick off. Optical waveguides (WG) represent a very attractive platform for the exploitation of such coherent effects in actual devices, due to an easy technological fabrication, the high quality factors achievable and the particular geometry, suitable for the integration of polaritonic optical circuits and coherent optical sources. A guided electromagnetic mode, localized in the two dimensional plane of the WG by total internal reflection, couples to an exciton confined to the quantum well (QW), giving rise to WG polariton modes \cite{Rosenberg2016, Walker2013}, in counterposition to the more reknown microcavity polariton modes \cite{Kavokin2008}. Recently, new WG designs have enabled the application of a transverse electric field, while keeping the system in the strong coupling regime \cite{Rosenberg2016, Rosenberg2018}. In this work, we indeed show that, despite the lack of an energy minimum in the WG polariton dispersion, an electrically controlled polariton lasing effect can be achieved with guided polaritons propagating at high speed thanks to a spatially localized population inversion. It is worth nothing that here we use the terminology "polariton lasing" to distinguish this effect – a mixture between lasing under weak coupling and injection into a polariton mode – from the phenomenon of polariton condensation.

We use two metal gratings placed on top of a GaAS/AlGaAs slab, that behave like a couple of semi-reflective mirrors, confining the WG polariton mode and outcoupling the laser emission vertically throughout one of their diffraction orders. Their design favors the formation of an energy gap in the energy momentum dispersion of the planar WG mode. Inside this energy interval, a manifold of Fabry-Perot (FP) modes is formed by the cavity effect between the gratings, funneling the light absorbed and reemitted by carrier recombinations under the pump spot. The lasing effect is then enabled on a specific polariton mode, i.e., the first mode for which gain equals losses upon increasing pump power. Parameters such as grating periodicity and filling factor as well as the cavity length (i.e.~distance between gratings) provide a fine control of the laser properties, without the need for complex postprocessing of the WG. Most importantly, using an electric field applied in the direction perpendicular to the WG plane, we demonstrate real-time tunability of the emission wavelength. We stress that this electrical control is possible only because of the WG geometry, that allows to place the electrodes in close vicinity to the quantum wells, as it has been previously reported on the same structure \cite{Rosenberg2016, Rosenberg2018, liran_fully_2018}. The strong coupling regime and the sensitivity of the exciton to the externally applied electric field allow us to obtain an electrically controlled polariton laser. These findings could be an important tool for the future development of polariton circuits, enabling the realization of a coherent source of propagating polaritons (if only a fraction of the polaritons in the lasing mode is extracted by the gratings), Q-switching, electro-optic modulators and configurable logic gates.

\section{Results}
\begin{figure}
    \centering
    \includegraphics[width=\columnwidth]{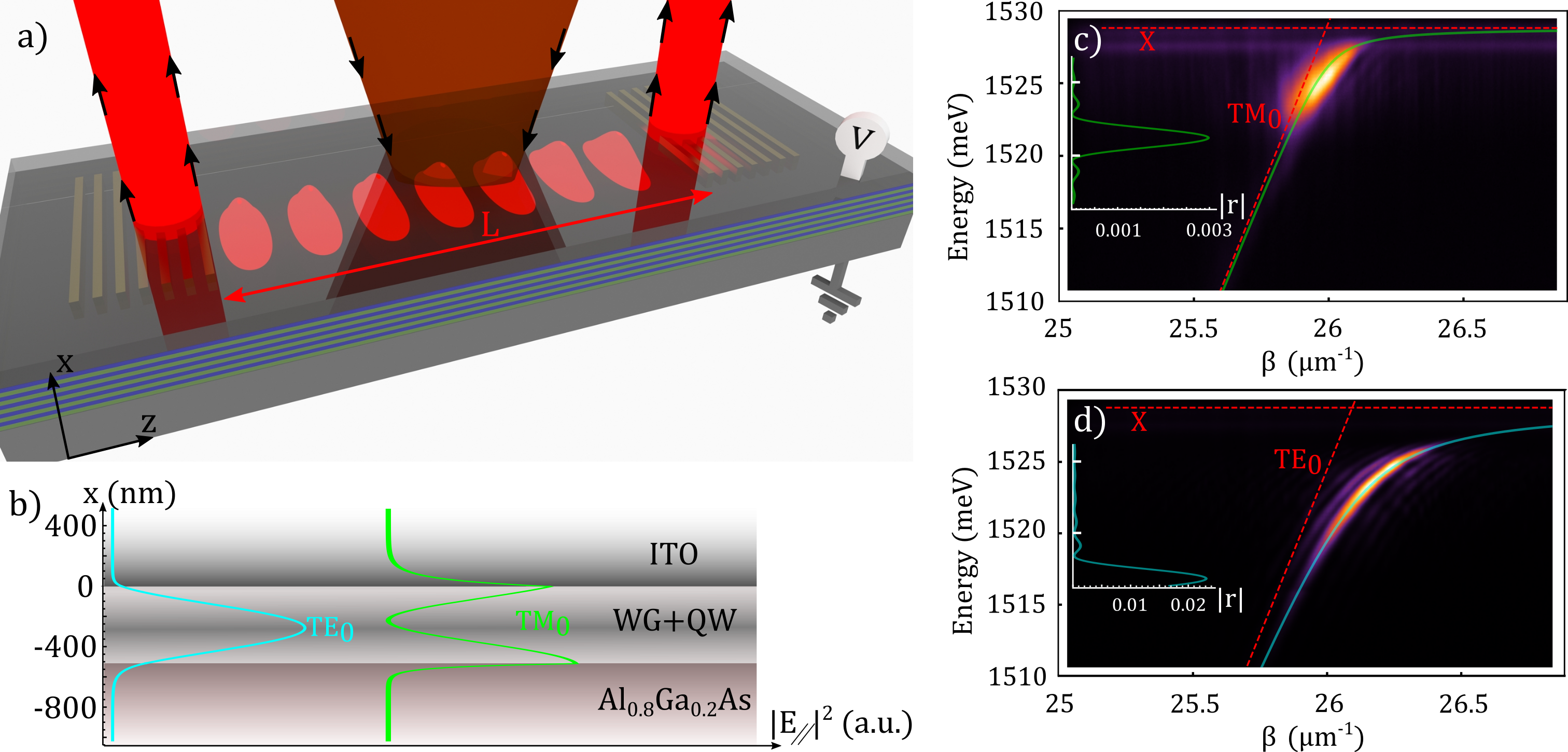}
    \caption{\textbf{a} Schematic representation of the system. Two gold gratings, placed on top of the WG, couple to the vanishing tail of the guided modes. As a result, a FP cavity is formed inside the slab for the TM mode, opening the possibility to observe a lasing effect in the system. L is the cavity length given by the distance between the gratings. The doped substrate, together with a top layer of ITO, enables the application of an electric field in the direction perpendicular to the cavity plane. \textbf{b} Calculated intensity profile of the in-plane electric field for each guided mode. The x component of the electric field for the TM$_0$ mode is not included in the scheme because it does not contribute to the light-matter coupling.
    \textbf{c-d} Polarization resolved PL dispersion of the system extracted through a single gold grating placed on top of the slab WG. Each dispersion is plotted as a function of the wavevector of the guided modes $\beta$. Panel c corresponds to TM polarization, while panel d shows the dispersion for the TE mode. Continuous lines correspond to a fitting of each mode by using a theoretical model of coupled oscillators. The obtained Rabi splitting ($\Omega$) values are $5.2$ meV for the TM mode and $13.4$ meV for the TE. The bare exciton and photonic modes are indicated by red dashed lines. The calculated reflectivity of the mode in presence of the grating is displayed as an inset for each polarization. The creation of reflectivity maxima indicates the opening of energy gaps.}
    \label{sketch}
\end{figure}

The WG structure is grown on an $n^+$-doped GaAs substrate, on top of which a $500$ nm cladding of $Al_{0.8}Ga_{0.2}As$ was previously deposited \cite{Rosenberg2016, Rosenberg2018, liran_fully_2018}. The structure consists of 12 pairs of $20$ nm thick GaAs quantum wells (QW) separated by $20$ nm of $Al_{0.4}Ga_{0.6}As$ barriers. Light is extracted through gold gratings spaced by a distance ranging from $50~\mu m$ to $150~\mu m$. To realize FP resonators inside the WG, two identical gratings are fabricated facing each other on top of the slab at a given distance along their line of sight (a sketch of the full system is shown in Fig.~\ref{sketch}a and the calculated spatial distribution of the electric field in-plane component is depicted in Fig.~\ref{sketch}b). We use gratings with pitch $\sim 240$ nm and filling factor comprised between $\sim0.72$ and $\sim0.85$. 
Finally, a $50$ nm layer of ITO is sputtered on top of the sample. This layer, together with the doped substrate in the opposite side, allows for the application of an electric field in a direction perpendicular to the slab plane. The purpose of the electric field is to tune the exciton energy, exploiting the Stark effect, and hence its influence on the polaritonic guided modes \cite{Rosenberg2016,Rosenberg2018,liran_fully_2018}. A more detailed description of the structure is given in the Materials and Methods section. 

To firstly verify the presence and features of the WG modes inside the slab, we use a nonresonant pump laser which excites the sample outside of the gratings region, while the photoluminescence (PL) spectrum of the system is collected through an individual grating (so no cavity is formed) placed $100$ $\mu m$ away from the off-resonant excitation spot. Figures \ref{sketch}c-d show the PL emission coming from the hybridized zero-th order Tranverse Magnetic (TM) and Transverse Electric (TE) modes of the structure, respectively. The dispersions are plotted as a function of the wavevector $\beta$ of the propagating mode inside the slab. The uncoupled bare exciton line (heavy hole exciton) and guided optical modes are indicated by horizontal and oblique red dashed lines, respectively. The experimental polariton dispersion can be fitted by means of a theoretical model of coupled oscillators, as shown by the solid lines in Fig.\ref{sketch}c-d. The results show an asymmetry in the Rabi splitting ($\Omega$) between the TE and TM modes. We obtain values of $13.4$ meV and $5.2$ meV, respectively. As a matter of fact this is predicted by the selection rules for the coupling of the confined modes with the exciton dipole, which imply an $\Omega$ value around three times higher for the TE mode than for the TM mode, as it has already been observed in similar samples \cite{Rosenberg2016,Shapochkin2018}. The figure also evidences an important difference in the resolution in $\beta$ with which the TE and TM modes are outcoupled. This linewidth effect is originated by the different grating efficiency with respect to each polarization and it is correlated with the formation of the FP cavity inside the structure, as will be described in detail in the Discussion Section.

\begin{figure}
    \centering
    \includegraphics[width=0.6\columnwidth]{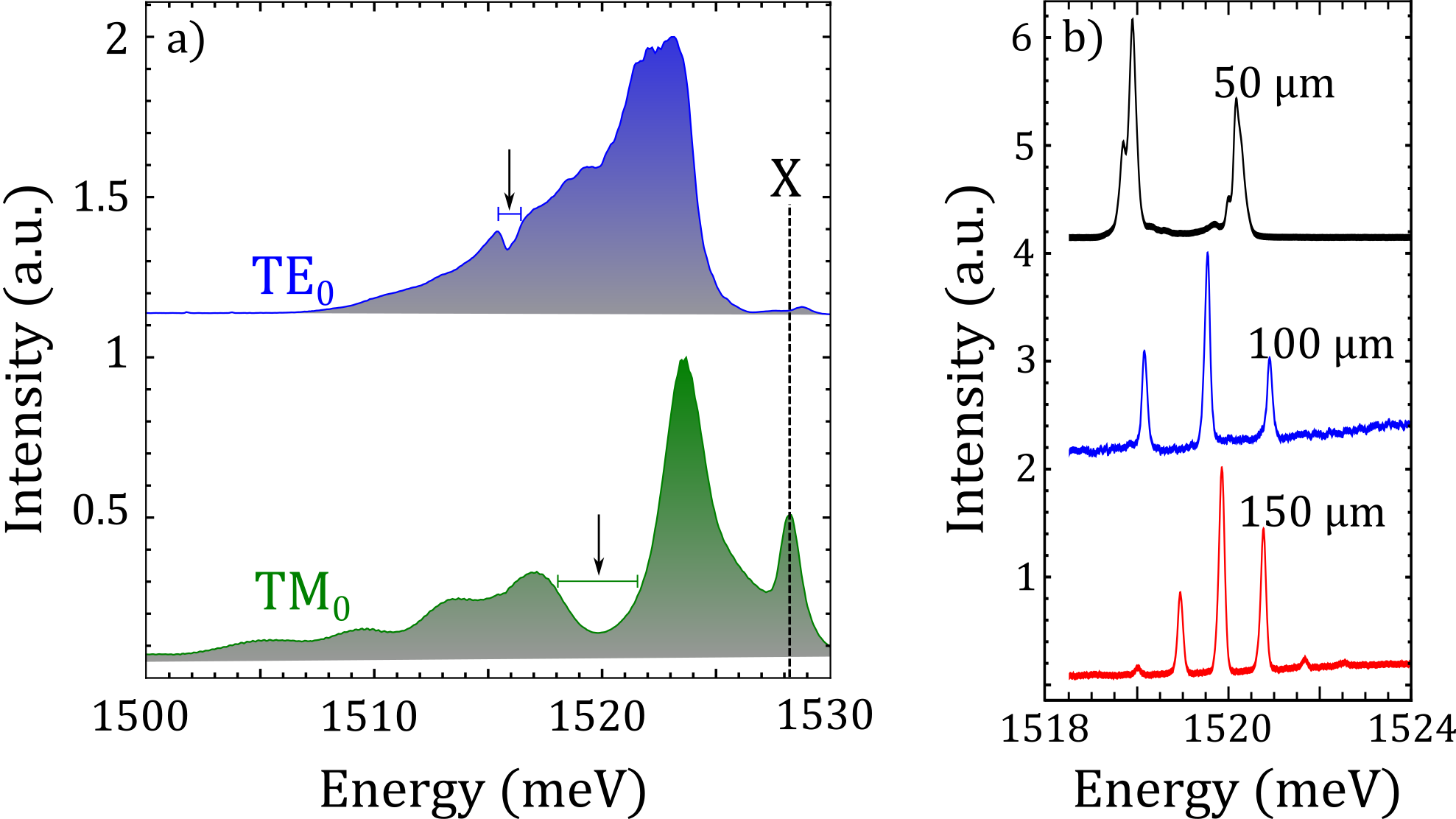}
    \caption{\textbf{a} Experimental measurement of the gratings gap in TE and TM polarization. The spectra are obtained by integrating the PL emission from the terminal end of a 400 $\mu$m-long gold grating over the emission angle. The vertical dashed line indicates the exciton energy. The opening of two bandgaps is highlighted by the vertical arrows. \textbf{b} Fabry-Perot modes formed along the TM$_0$ dispersion. The constructed FP cavities have different lengths, determined by the distance between the gratings. As expected for an optical cavity, the smaller the cavity, the greater the free spectral range at a given energy. The spectral distances of $2.3$ meV for the cavity of length $50$ $\mu$m, $1.4$ meV for the cavity of $100$ $\mu$m and $0.9$ meV for the cavity of $150$ $\mu$m, are associated to FP modes of manifold $17$, $30$ and $45$, respectively. In the three cases, the system is pumped near the threshold power (110 mW, 110 mW and 150 mW, respectively.), where the FP modes become visible, and the spot is placed at the center of the FP cavity.}
    \label{gaps}
\end{figure}
After the WG modes, we also characterize the features of the top gratings and their role in forming the the FP cavity. Figure \ref{gaps}a shows the angle-integrated PL emission intensity from the terminal end of a single and very extended, 400 $\mu$m-long grating as a function of the emission energy. Both the TE and TM polarization possess a local minimum in the emitted intensities, highlighted by vertical arrows in the figure. These minima represent energy gaps corresponding to wavevectors of the guided modes that are back-reflected by the grating. In other words, each grating acts as a mirror in the energy interval corresponding to the gaps of Fig.~\ref{gaps}a and, when two gratings face each other, a planar FP cavity is formed, with a cavity length which depends on the distance $L$ between gratings. The cavity modes are characterized now thanks to the PL emission outcoupled from one of the gratings when a nonresonant pumping spot is placed in the middle between the two (as illustrated in Fig.~\ref{sketch}a). Figure \ref{gaps}b shows the PL emission of the cavity modes formed along the TM$_0$ mode by placing the gratings at distances of $L=50$ $\mu$m (black upper line), $L=100$ $\mu$m (middle blue line) and $L=150$ $\mu$m (lower red line). As expected, the closer the gratings to each other, the larger the free spectral range. In the case of the gratings separated by $50$ $\mu$m the modes spacing is $\sim2.3$ meV that corresponds to the FP mode of manifold $17$. For the case of a pair of gratings separated $100$ $\mu$m, the spacing is $\sim1.4$ meV (manifold 30) and for the gratings separated $150$ $\mu$m the spacing is $\sim0.9$ meV (manifold 45). Our system then, is not operating near the global energy minimum, which would occur at the smallest guided wavevector (i.e. at a much smaller energy and for a negligible excitonic fraction). It operates instead at the local minima of the FP modes, i.e., at the bottom of the parabolic dispersion of each FP mode.
The observed FP modes are then an effect of the confinement of the TM$_0$ mode due to the presence of two identical gratings.
A comparative analysis of the extracted PL is performed with respect to the fabrication and materials features of the grating(s) (Fig.~S4) and shows that when the metallic grating is in direct contact with the dielectric guide it carries a considerable spectral broadening of the measured TM mode (and not of the TE one). We assign this broadening to an additional residual absorption from plasmonic modes in the gold grating, a widely studied effect \cite{Klein2005,Zhang2012,Gollmer2017}. 
It is important to note that this possible plasmon induced absorption takes place only in the grating region and does not modify the truly waveguide polariton character of the propagating modes outside the gratings, as demonstrated in Fig.~S4c, where the PL emission collected through a fully dielectric grating confirms the expected behavior for a system operating in the strong coupling regime. For an extended discussion with respect to the grating features, we refer the reader to the SM.

\begin{figure}
    \centering
    \includegraphics[width=0.8\columnwidth]{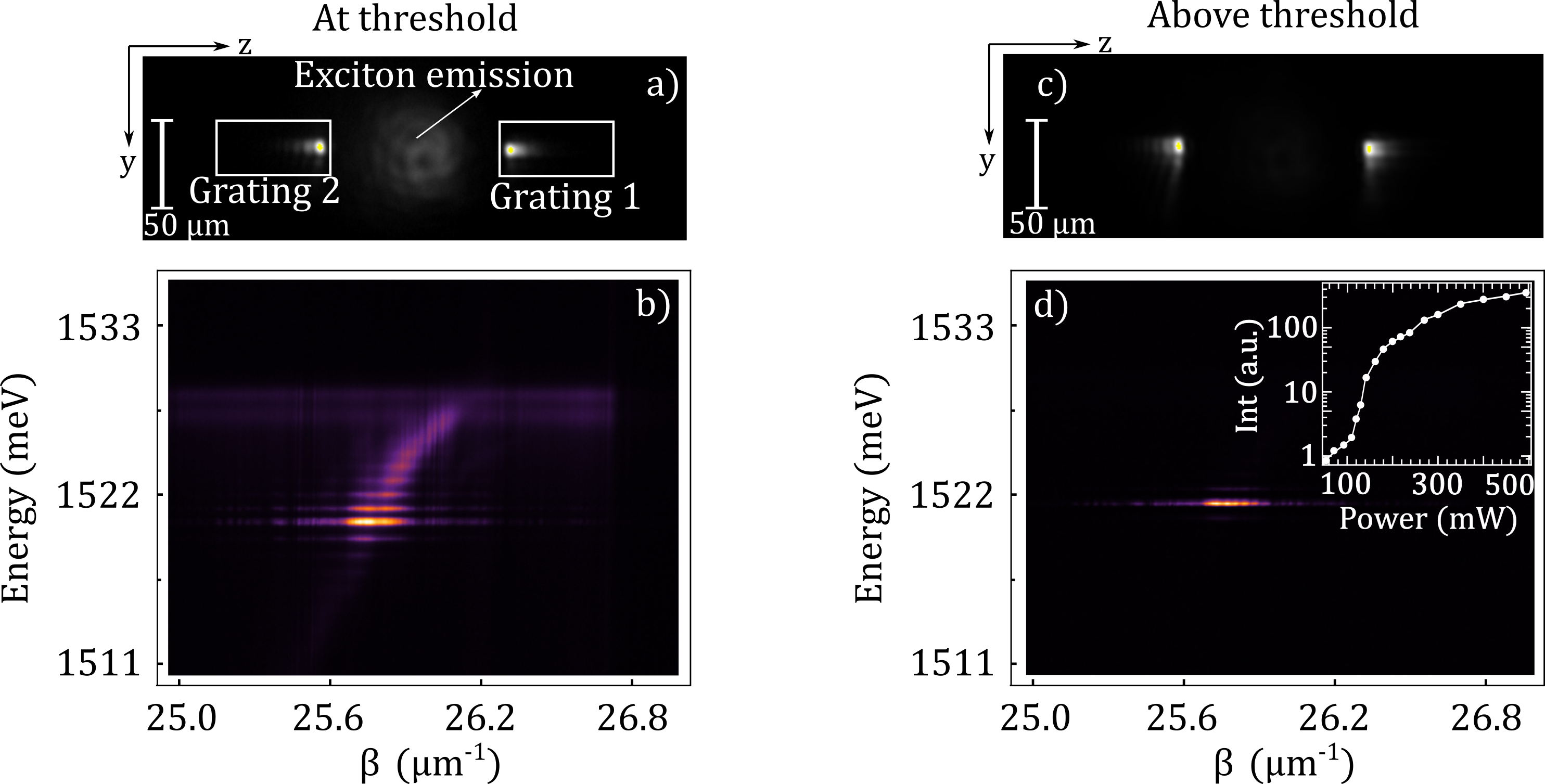}
    \caption{\textbf{a} Real space image of the 100 $\mu m$ long FP cavity when excited out of resonance near the threshold power (95 mW). The laser spot is placed in the middle of the cavity, exciting both, the QW exciton with low in-plane momentum (central spot) and the polaritonic guided modes, extracted from the gratings (indicated with white rectangles). The residual pump laser signal is suppressed with a long-pass spectral filter. \textbf{b} The TM mode is reconstructed in the Fourier space by spatially isolating the PL from grating 1. The insertion of the gratings entails an additional confinement, and hence, the formation of discrete modes along the TM dispersion, as it can be observed in the image. \textbf{c} When the pump power overcomes the threshold (140 mW), one mode of the cavity gets massively populated, reaching an emission intensity much higher than any other emission in the system. \textbf{d} the PL dispersion measurement (measured as in b) confirms which state of the FP cavity gets populated. The inset in \textbf{d} shows the intensity of the lasing mode as a function of the pumping power; a clear threshold behaviour is visible.}
    \label{threshlod}
\end{figure}

We have characterized all the three building blocks (namely the WG modes, the gratings and the planar FP cavity), which are mandatory for the observation of the waveguiding lasing effect. We can hence shine a high peak-power nonresonant laser in a region between the two gratings, in the specific we use a $100$ fs pulsed laser tuned at $1.59$ eV ($\approx780$ nm) and 80 MHz repetition rate, with a spot size of $40$ $\mu$m. Figure \ref{threshlod}a shows the PL emission in real space at threshold excitation power. The big spot between the gratings corresponds to the exciton emission under the pump spot which is not coupled to the guided modes. The excited guided polariton modes propagate along the slab with in-plane vector $\beta$ and are partly extracted and partly reflected back in the plane when they hit the gratings, highlighted by white rectangles. The PL emission extracted from grating 1 can be resolved in both energy and momentum (Fig.~\ref{threshlod}b). Remarkably, such emission shows a series of discrete peaks along the TM polariton dispersion corresponding to the FP modes of Fig.~\ref{gaps}b. In other words, in the region of space comprised between the two identical gratings, a FP cavity is formed, sustaining the discrete modes observed in Fig.~\ref{threshlod}b. With increasing pump power, as shown in Fig.~\ref{threshlod}c-d, the system shows a clear threshold behaviour (see inset of panel d) with prominent coherent emission from only one of the FP modes at an energy of $\sim 1520.5$ meV, that in the dispersion relation corresponds to a wavevector $\beta=25.96$ $\mu m^{-1}$. This is the precise in-plane wavelength corresponding to the grating pitch $\Lambda$, where it is hence maximally reflective. A strong reduction of the FP mode linewidth is observed across the lasing threshold (Fig.~S5a). Moreover, when the emitted light from the two gratings is overlapped in $k$ space, an interference pattern is generated (see Figs.~S5b to g for the build-up of the first order coherence across the lasing threshold). The threshold and linewidth effects are clear signatures which mark the onset of the lasing regime in the system.

\begin{figure}
    \centering
    \includegraphics[width=0.8\columnwidth]{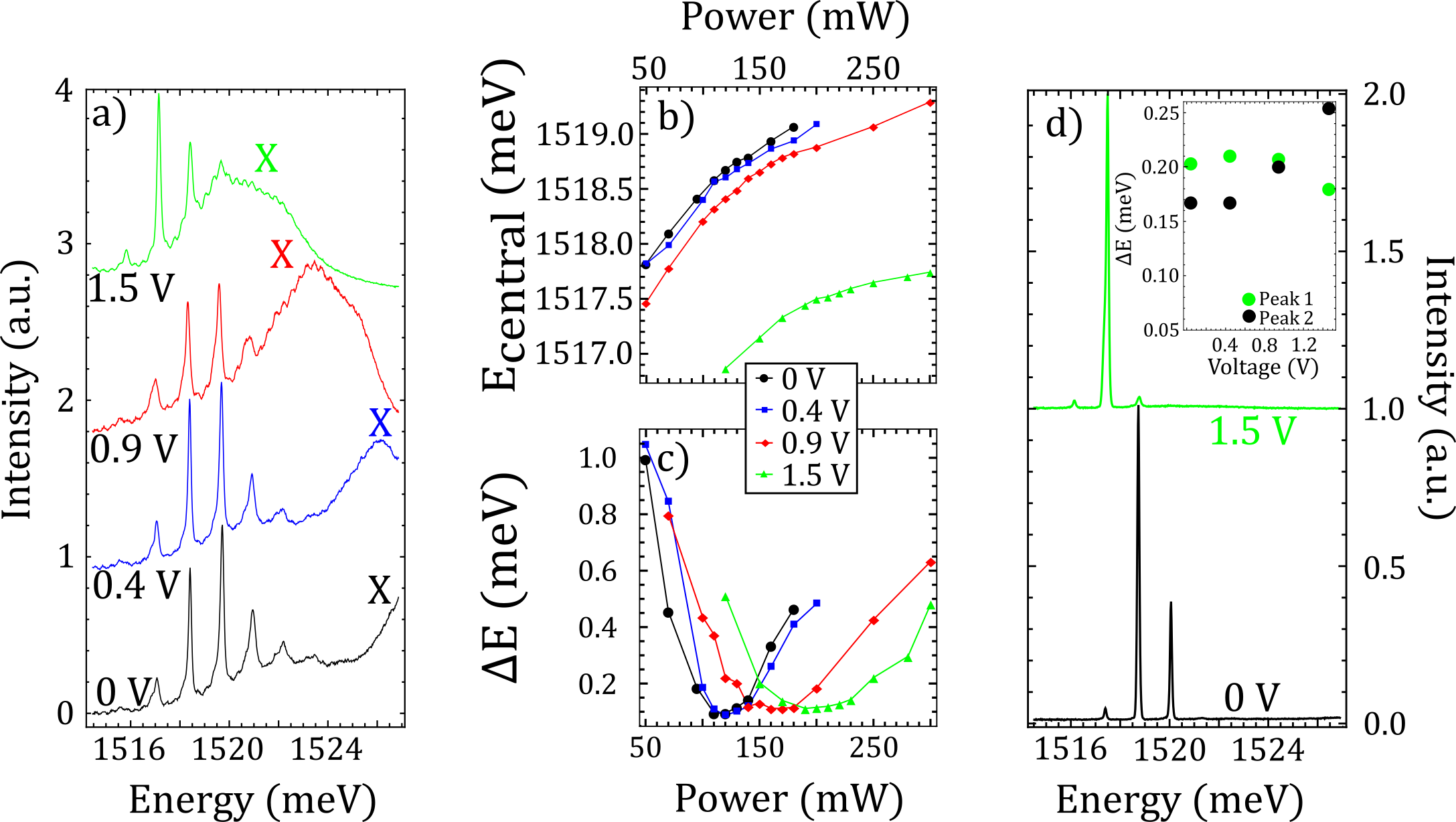}
    \caption{\textbf{a} Spectra obtained from a profile of the TM polariton dispersion at $\beta \approx 26$ $\mu m^{-1}$, and for different applied voltages. The letter X indicates the peak corresponding to the PL emission of the uncoupled exciton that redshifts due to the Stark effect. \textbf{b}-\textbf{c} To verify that the presence of a transverse electric field does not affect the properties of the lasing effect, we measure the central energy shift (b) and spectral narrowing (c) for three applied voltages. The results not only confirm the lasing effect in every case, but show that the electric field acts as a fine tuning of the laser energy for small voltages, and as a switch that allows to select the cavity mode in which the lasing takes place. The latter effect is achieved as a consequence of a modification of the relative Q factor of the cavity modes, since the exciton resonance affects the absorption at the energy of each FP mode. \textbf{d} Spectra of the cavity taken at threshold power without any transverse electric field (black) and with an applied voltage of $1.5$ V (blue). The modes energy is almost unaltered, but the change of their absorption properties switches the mode in which the stimulated emission takes place. The inset shows the FWHM of each peak as the applied electric field increases. When the quality factor of peak 1 decreases below the one of peak 2, the lasing mode switches. The data is acquired in the FP cavity of $100$ $\mu m$ while pumped at the center. The emission is collected from one of the gratings in the far-field.}
    \label{voltage}
\end{figure}

Finally, we can now apply an external electric field perpendicular to the plane of the slab, demonstrating an extra degree of freedom acting as an additional control parameter on the WG polariton lasing. Figure \ref{voltage}a shows the spectra obtained from the TM polariton dispersion at $\beta \approx 26~\mu m^{-1}$ and for different applied voltages. The TM dispersion is modified due to the Stark effect that red-shifts the exciton energy. In Fig.~\ref{voltage}b-c we measure the central emission energy and the linewidth of the lasing mode for three different applied voltages and as a function of the pump power. For small applied fields ($\leq 0.9$ V), a red-shift of the lasing mode is observed. When the applied electric field is further increased ($>0.9$ V), the exciton energy approaches the cavity modes, more largely altering their absorption properties, and hence, their relative losses. As a matter of fact, Fig.~\ref{voltage}d shows that for an applied voltage of $1.5$ V, the mode in which the stimulated emission takes place jumps to the next manifold (both spectra are taken at threshold power). The electric field acts in this case as a neat switch that allows to select the mode in which the lasing takes place by modifying the FP modes relative losses. The inset of Fig.~\ref{voltage}d shows the evolution of the linewidths of the two lasing peaks upon application of the electric field. Indeed the switch between the lasing modes also coincides with the crossing of their linewidths. 
We can access the temporal dynamics of the process by means of a streak camera coupled to a monochromator and a scan in the reciprocal space. The energy-momentum dispersion resolved at different times is shown in Fig.~\ref{frames} (see also Supplementary Video 1 and 2). In the case of Figs.~\ref{frames}a-c the pump spot is placed between the two gratings and the emission from only one of the gratings is recorded (see Fig.~\ref{frames}d). Figure \ref{frames}a shows the far-field emission 200 ps after the arrival of the pump pulse (at the lasing threshold power). The FP modes are clearly visible. These modes are in TM polarization, i.e., the polarization for which the two gold gratings act as mirrors. Even when the FP modes population fades away, Figs.~\ref{frames}b-c, the TE mode (labelled "TE Downward") is still visible. Figures \ref{frames}e-g show the temporal dynamics when the pumping spot is placed on one of the two gratings, G1, and the far-field emission is collected from the whole cavity area (see sketch in Fig.~\ref{frames}h). Again, after about 200 ps from the arrival of the pump pulse, we observe lasing emission from the TM modes, see Fig.~\ref{frames}e (note the interference coming from the superposition of the light emitted from G1 and G2). Interestingly, the portion of the sample under the pumping spot (grating G1) shows both weak and strong coupling features (respectively dotted and dashed dispersions in Figs.~\ref{frames} e-f), while the light coming from the grating G2 shows strong coupling features (e.g.~the TE mode propagating downward). Figures \ref{frames}f-g show that as carriers recombine and weak coupling features are lost, the lasing emission fades away. These observation strongly suggests a connection between weak coupling under the pumping spot and the polariton lasing emission. We also note that in Figs.~\ref{frames}e-g the TE mode propagating upward is not visible. We think that this is due to the fact that only a small portion of the grating G1 is free from the pumping spot and able to extract the TE mode propagating upward (while the whole G2 contributes to the extraction of the downward propagating TE mode). Moreover, the extraction efficiency of the gratings is smaller for the TE modes than for the TM modes, which explains why the upward propagating TM mode is instead visible. 

\section{Discussion}
We demonstrated the design and realisation of a micrometer size laser in a GaAs/AlGaAs polariton-WG by using two metallic gratings at controlled distance on top of the slab WG, in order to form a FP cavity that confines the photonic field in one dimension. To better understand this effect, we calculate the influence of the grating on the guided modes by using a perturbation formalism to study the coupling between counter-propagating modes \cite{Stoll1973,Yariv1973}. For a WG grown in the $x$ direction, and with modes propagating in the $z$ direction, the magnetic  field $H(x,z,t)$ of the $m^{th}$ TM mode, can be expressed in case of an infinite slab in the $y$ direction (i.~e.~$dH/dy=0$) as:
\begin{equation}H_m(x,z,t)=\mathcal{H}^m_y(x)e^{i(\omega t-\beta_m z)}\end{equation}
Where $\mathcal{H}(x)$ is derived from the corresponding wave equation and boundary conditions under the ``long wavelength" approximation, that takes into account the internal WG periodicity (12 pairs of QW+barrier) by considering an effective refractive index. A numerical verification of the validity of this approach, upon comparison with a full transfer matrix method, is presented and discussed in Fig.~S2 of the SM, together with its detailed explanation, based on ref.~\cite{Yariv2007}. The fabrication of a grating with periodicity $\Lambda$ on top of the WG acts as a perturbation that changes the effective in-plane momentum of the $m^{th}$ mode as:
\begin{equation}
    \beta'_m=\frac{l \pi}{\Lambda}-\sqrt{\left(\beta_m-\frac{l \pi}{\Lambda}\right)^2-|\kappa|^2}
\end{equation}
With $\beta_m$ the in-plane momentum of the unperturbed $m^{th}$ mode, $l$ an integer number such as $|\beta_m-\frac{l \pi}{\Lambda}|\sim0$ and $\kappa$ a constant that depends on the grating periodicity and material, and the spatial profile of the transverse field. In the case of a square grating fabricated on top of the slab, with filling factor $a$, height $h$ and refractive index $n_g$, the constant $\kappa$ takes the form
\begin{equation}
\label{ktm}
\kappa=\frac{i \omega \mu_o}{4\pi l} Sin\left(\pi l a\right)\int_{0}^{h}\left(\left(n_g/n_{ITO}\right)^2-1\right)\left[\mathcal{H}_y^m(x)\right]^2 dx
\end{equation}
This expression indicates the existence of an energy gap in the guided mode dispersion in the region covered by the grating when $|\kappa|^2>\left(\beta_m-\frac{l \pi}{\Lambda}\right)^2$. At the energy interval around the exciton ($\sim 1.529$ meV), where the electromagnetic field hybridizes, the in-plane linear momentum of the TM guided mode lies in the range ($25.6$ $\mu m^{-1}<\beta_m<26.3$ $\mu m^{-1}$), as observed in Fig.~\ref{sketch}c. For a grating with period $\Lambda=243$ nm, the condition $\beta_m-\frac{l \pi}{\Lambda}\sim0$ is satisfied for $l=2$. The expression \ref{ktm} shows that, for $l$ even, $\kappa$ differs from zero only for filling factors different than $0.5$. Moreover it is also evident from the expressions that for $l=2$ the maximum gap's amplitude is obtained for a filling factor either $0.25$ or $0.75$. In our case, we fabricated gold gratings with filling factor close to $0.75$ to maximize the coupling between propagating and counter-propagating modes and hence their reflectivity, whose calculation is shown on the left side of Figs.~\ref{sketch}c-d. The Purcell enhancement effect induced by the semi-reflective gratings provides the conditions for the lasing from the FP mode with energy $E\sim 1520.5$ meV and in-plane wavevector $\beta\sim 25.96$ $\mu m^{-1}$ (Fig.~\ref{threshlod}d). The theoretical dispersion fitting reveals that the excitonic fraction of the coherent polariton population is $|C_x|^2=0.1$ at 0V. This value increases for an applied electric field, given that it only slightly modifies the position of the FP modes while it strongly affects the polaritonic dispersion. In other words, while the excitonic resonance explicitly modifies the polariton dispersion, its tail produces only small changes in the FP modes. An attractive feature of this system is that the excitonic component of the generated coherent population can be tailored by engineering the parameters of the FP cavity; in particular its length (distance between gratings) and the grating period. Hence, no modification of the slab WG is required.

\begin{figure}
    \centering
    \includegraphics[width=\columnwidth]{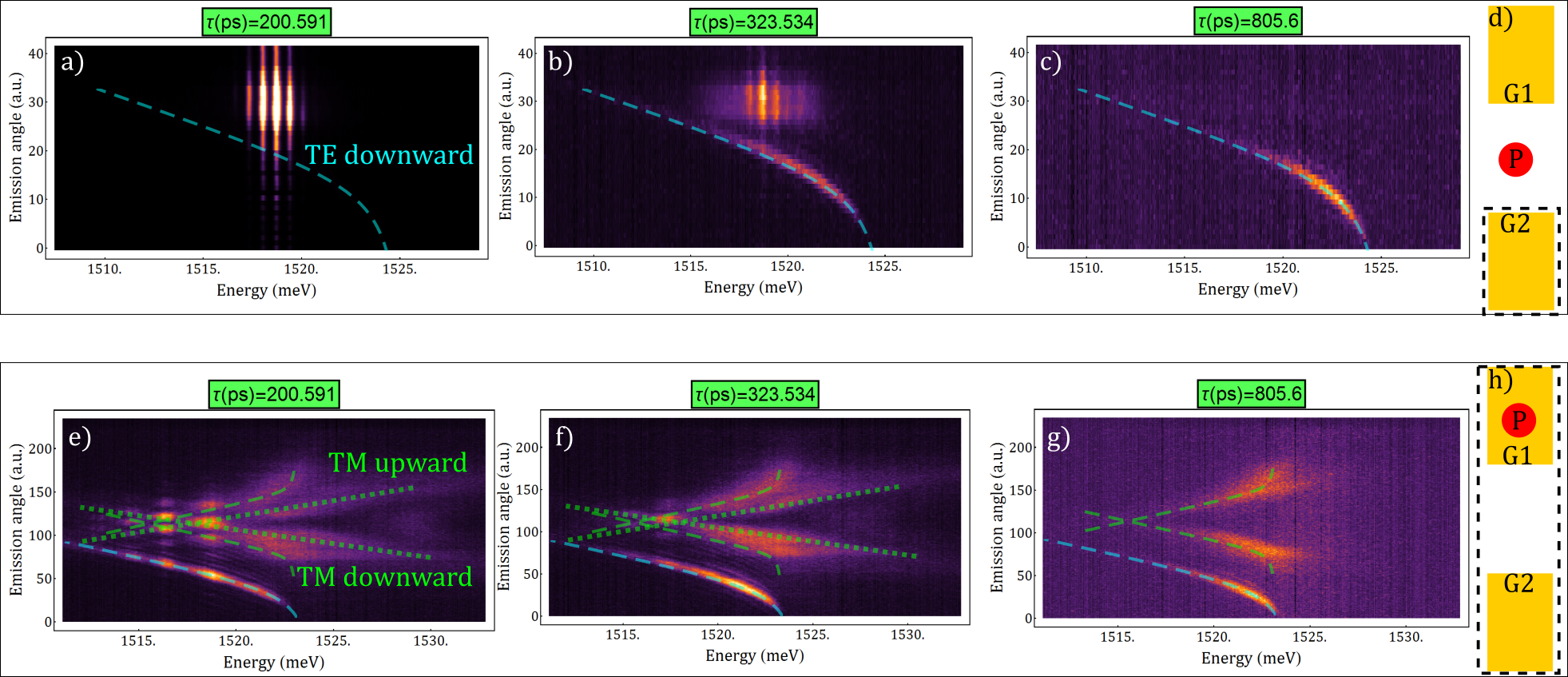}
    \caption{Temporal dynamics of the system after the arrival of the pumping pulse. The upper panels correspond to the case in which the FP cavity is pumped at the center and the emission is collected from the lower grating (see sketch in panel d), while in the lower panels the pumping laser is placed on the upper grating and the emission is collected from both gratings and overlapped in the Fourier plane (see sketch in panel h). In each case there is a visual guidance to identify the TE mode in the upper panel, and the coupled TE (dashed blue), TM (dashed green) and uncoupled TM (dotted green), in the lower panel. The cavity has a length of $100$ $\mu m$ and is pumped with a power of 95 mW. In panels d and h the red dot indicates the pump spot position and the yellow rectangles the metallic gratings. The dashed black line highlights the spatial region from which the emission is collected.}
    \label{frames}
\end{figure}

It is worth noting that the blueshift of Fig.~\ref{voltage}b cannot be simply associated to polariton-polariton interactions, since the exciton energy remains almost unaltered when the pump power is increased. It rather might depend on a local carrier-induced reduction of the WG refractive index \cite{44924}, due, for example, to the plasma generated by the optical pump pulse, i.~e., a plasma induced transparency \cite{Chow1988,Murata1993} or to the band filling effect. Such effects can be especially relevant in the spatial region below the pump spot. An estimation of the index change with increasing pump power can be done by considering that the $i^{th}$ mode of the resonator has energy $E_i=i\frac{h\cdot c}{2 L n}$, where $L$ is the cavity length, given by the distance between the gratings, $n$ its effective refractive index, $h$ the Planck's constant and $c$ the vacuum light speed. The data shows a saturation behavior compatible with such explanation, with a reduction of only $10^{-4}$ in the refractive index. Figure \ref{voltage}c shows the linewidth for different pump powers and voltages. The observed spectral narrowing at threshold is expected and it is associated to the fact that the coherence time of the system exceeds the radiative lifetime of the confined mode under these excitation conditions \cite{Bajoni2008}. Based on the optical characterization of the system and the complex temporal dynamics that we observe in Fig.~\ref{frames} (see also Supplementary Video 1 and 2) we can explain the laser build-up in the following way: the gain media (QWs) are populated through the non-resonant pump in the middle of the FP cavity, as shown in Fig.~\ref{threshlod}a, forming an electronic plasma localized under the pumping spot. After the recombination of photo-created carriers, the photons can be emitted in any of these FP modes (as the Purcell factor enhances the emission in them). By following the standard laser theory \cite{Svelto2010} the first mode to start to lase is the one for which a threshold condition is met. In the dynamics of our system, this condition is achieved after about 200 ps, as shown in the panels a and e of Fig.~\ref{frames}. This condition is written in terms of the critical population inversion: $N_C=\gamma/\sigma L$; where $N_C$ is the population at which the lasing starts and $\sigma$ is the spontaneous emission coefficient. Here $\gamma$ represents the overall losses of a given mode as $\gamma=-ln(R_1)-ln(R_2)-ln(1-L)$ in which $R_1$ and $R_2$ are the mirror reflectivity and $L$ represents the losses for a roundtrip in the cavity. The first mode to start to lase is then the mode with the smaller $N_C$, i.e., the mode with the smaller linewidth in the set of the FP modes created by the two gratings. By modifying the linewidth of the modes, as it is shown in Fig.~\ref{voltage}d, it is possible to change the mode which starts to lase. It is important to remark that while the mechanism of the laser can be simply explained, the coherent population generated has an interactive nature, which could carry some complex effects of competition between modes, perhaps involving polariton nonlinearities. While these effect could be interesting they are beyond the scope of the present work. An additional feature revealed by Fig.~\ref{frames}e-h is the particular coexistence between populations in strong and weak coupling inside the FP cavity, see also Supplementary Video 2. The signal extracted from the upper grating (G1), shows a superposition of population in strong coupling (green dashed lines) and weak coupling (green dotted line). This population has a much longer lifetime than the FP modes, as it can be observed in panel f, where the coherent population has already left the cavity, but the coupled and uncoupled TM population is still present. While we detect complex emission patterns under the laser spot, that can be identified as carrier recombination in weak coupling regime, the emission from G2 corresponds to a WG polaritonic dispersion in strong coupling. This evidence goes in the direction of the creation of an electronic plasma under the pump spot, i.e., to an inversion of population. However, this plasma is restricted exclusively to the illuminated region (since the excitonic group velocity is too low to provide significant propagation during the excitonic lifetime). The radiative decay of this population into the polaritonic guided mode of the FP cavity with the lowest losses enables the massive population of such state, creating the coherent state of guided polaritons. The full process can then be interpreted as the coexistence of an electronic gas (in weak coupling) that preferably decays into a polaritonic guided mode, generating a coherent population of propagating hybrid light-matter particles (in strong coupling). The long lived exciton population keeps feeding the cavity modes during the full lasing process and after the emission of the coherent population, as observed in panel c and g, a proof that the quality factor of the slab waveguide is much higher than the one of the FP cavity. The observation of the coupled TE mode (dashed blue line) during the full time range demonstrates the coexistence of dressed and bare states inside the cavity.
We note also that recently a polariton laser effect has been claimed at both cryogenic and room temperature in ZnO based WGs \cite{Jamadi2018}. In that case, the polariton emission has been interpreted as a standard polariton condensate without need for a population inversion. While this interpretation can be supported by the large Rabi splitting observed in ZnO systems, our present work shows that in guided polariton systems an alternative interpretation can explain the same features. Moreover the absence of a blueshift in case of Ref.~\cite{Jamadi2018} could be also compatible with a refractive index change in the region between the two cracks. A careful check of the dynamics both under and outside the pumping region is essential in this kind of systems to better clarify the lasing mechanism involved.

It is also worth noting that electrical control of polariton properties such as plasmon-exciton coupling \cite{Grossmann2011}, spin switch \cite{Dreismann2016} and parametric gain \cite{Christmann2010} have been reported. While the optimization and the precise characterization of the electrical control presented here is beyond the scope of this work, it is interesting to compare the theoretical minimum electric energy required for the switch of figure \ref{voltage}d with the value reported for the spin witch of \cite{Dreismann2016}. To do this we consider our heterostructure as a capacitor with capacity $C = 9*10^{-13} F/m$. In an optimized version of the present device, electrodes having an area as low as $100 \mu m*150 \mu m$ would be enough to apply the electric field of $1V$ required in figure \ref{voltage}d. Under these hypotheses, the required switching energy would be of the order of $600 fJ$. This value is about two order of magnitude higher than the value reported in \cite{Dreismann2016}, which is probably due to different physical mechanism involved: the spin state of the condensate is very sensitive to crystal birefringence and heavy-hole/light-hole mixing and a small applied electric field is enough to achieve the spin switching. In our case, the mode switching is obtained by shifting the exciton energy by Stark Effect, requiring higher applied electric fields and an increased amount of energy.

In summary, we report a lasing effect providing a coherent population of high-speed, guided polaritons. The additional optical confinement necessary to achieve the laser emission is provided by placing two specifically designed gold gratings on top of the WG. It is worth to note that this design is versatile, easy-to-fabricate and does not require complex post-processing of the WG, such as vertical selective etching. Most important, thanks to the WG geometry, we demonstrate an effective electrical control of the vertical laser emission. Remarkably, by electrically tuning the quality factor of the modes, we obtain an electrically switched polariton laser. This effect could be exploited for the realization of electrically Q-switched sources of polaritons. More generally, the lasing effect we report in this work would allow the construction of tunable, micrometer-size coherent sources of polaritons with different properties in a single wafer of slab WG. We think that our findings could be valuable in polaritonics especially in the development of polaritonic circuits and integrated polaritonic logic elements.

\section{Materials and Methods}

\textbf{Waveguide sample}
The full structure is grown on top of an n$^+$-doped GaAs substrate 500 $\mu$m thick. The cladding layer is made of $500$ nm of Al$_{0.8}$Ga$_{0.2}$As, while the WG is composed by $12$ bilayers of Al$_{0.4}$Ga$_{0.2}$As barrier ($20$ nm thick) and GaAs QW ($20$ nm thick), and a final bilayer made of $20$ nm of Al$_{0.4}$Ga$_{0.2}$As and $10$ nm of GaAs.\\

\textbf{Fabrication of gold gratings and deposition of ITO}
In order to fabricate gold gratings onto the sample we relied on a lift-off process. By design, they are $100$ $\mu m$ long, $50$ $\mu m$ wide and have a pitch of $243$ $nm$. Their production requires a positive e-beam resist PMMA A4 to be spun at $4000$ rpm onto the sample and baked for three minutes at $180$ $^\circ$C. The latter is then exposed using Raith150 and developed in MIBK:IPA 1:3. Subsequently, $3$ nm of chromium and $30$ nm of gold are thermally evaporated. The sample is placed in Remover PG at approximately $80$ $^\circ$C to remove the resist. A SEM characterization of the obtained gratings is displayed and discussed in the SM (Fig.~S3). The dense pattern carries irregularities in the first few grating wires related to the ``proximity effect" \cite{Kim2013}, very common in electron beam lithography. As shown and discussed in Fig.~S3, the affected area is negligible with respect to the full grating sizes. In closing, the electric contact is obtained by uniformly sputtering $50$ nm of Indium Tin Oxide (ITO) on top of the sample; the gold grating results completely covered in the end.\\

\textbf{Optical measurements}
For all the optical characterization the sample is kept at cryogenic temperature of $4$ K. All the PL measurements are realized in a confocal configuration, using a $100$ fs pulsed laser with a repetition rate of $80$ MHz, tuned at $1.59$ eV ($\approx780$ nm) to excite the sample out of resonance. The detection system allows to reconstruct either real or Fourier spaces in a Charge Coupled Device (CCD) coupled to a monochromator $70$ cm long with a diffractive grating with either $600$ or $1800$ lines per mm. This way it is possible to perform measurements resolved in space, angle and energy. An image of the real space is reconstructed before the CCD in order to apply a spatial filter by using a slit, enabling the selection of one or both gratings. The residual laser signal is suppressed with a long-pass spectral filter at $1.55$ eV ($\approx800$ nm). The time resolved images are performed in the same configuration, but directing the signal into a streak camera after passing the monochromator. The temporal reconstruction of the far field resolved in energy is made by moving vertically the focusing lens with a motorized station before it reaches the streak camera. Further details on the experimental setup and its schematic representation are presented in the SM.\\

\textbf{Acknowledgments}\\
We thank Paolo Cazzato for thechnical support.\\
We are grateful to R. Rapaport for inspiring discussions and for sharing information about the sample design.\\
Work at the Molecular Foundry was supported by the Office of Science, Office of Basic Energy Sciences, of the U.S. Department of Energy under Contract No. DE-AC02-05CH11231.\\
We thank Scott Dhuey at the Molecular Foundry for assistance with the electron beam lithography.\\
The authors are grateful to the - "Tecnopolo per la medicina di precisione" - (TecnoMed Puglia) - Regione Puglia: DGR n.2117 del
21/11/2018, CUP: B84I18000540002 and  "Tecnopolo di Nanotecnologia e Fotonica per la medicina di precisione" -(TECNOMED) - FISR/MIUR-CNR: delibera CIPE n.3449 del 7-08-2017, CUP: B83B17000010001.\\
The authors acknowledge the project PRIN Interacting Photons in Polariton Circuits – INPhoPOL (Ministry of University and Scientific Research (MIUR), $2017P9FJBS 001$).\\ 
This research is funded in part by the Gordon and Betty Moore Foundation's EPiQS Initiative, Grant GBMF9615 to L. N. Pfeiffer, and by the National Science Foundation MRSEC grant DMR 1420541.

\bibliographystyle{unsrt}
\bibliography{biblio.bib}

\newpage
\onecolumngrid

\setcounter{equation}{0}
\setcounter{figure}{0}
\setcounter{table}{0}
\setcounter{page}{1}
\makeatletter
\renewcommand{\theequation}{S\arabic{equation}}
\renewcommand{\thefigure}{S\arabic{figure}}
\pagenumbering{roman}

\noindent
\textbf{\Large Supporting Information}\\
\section{Experimental setup}
\begin{figure}[h!]
    \centering
    \includegraphics[width=0.65\columnwidth]{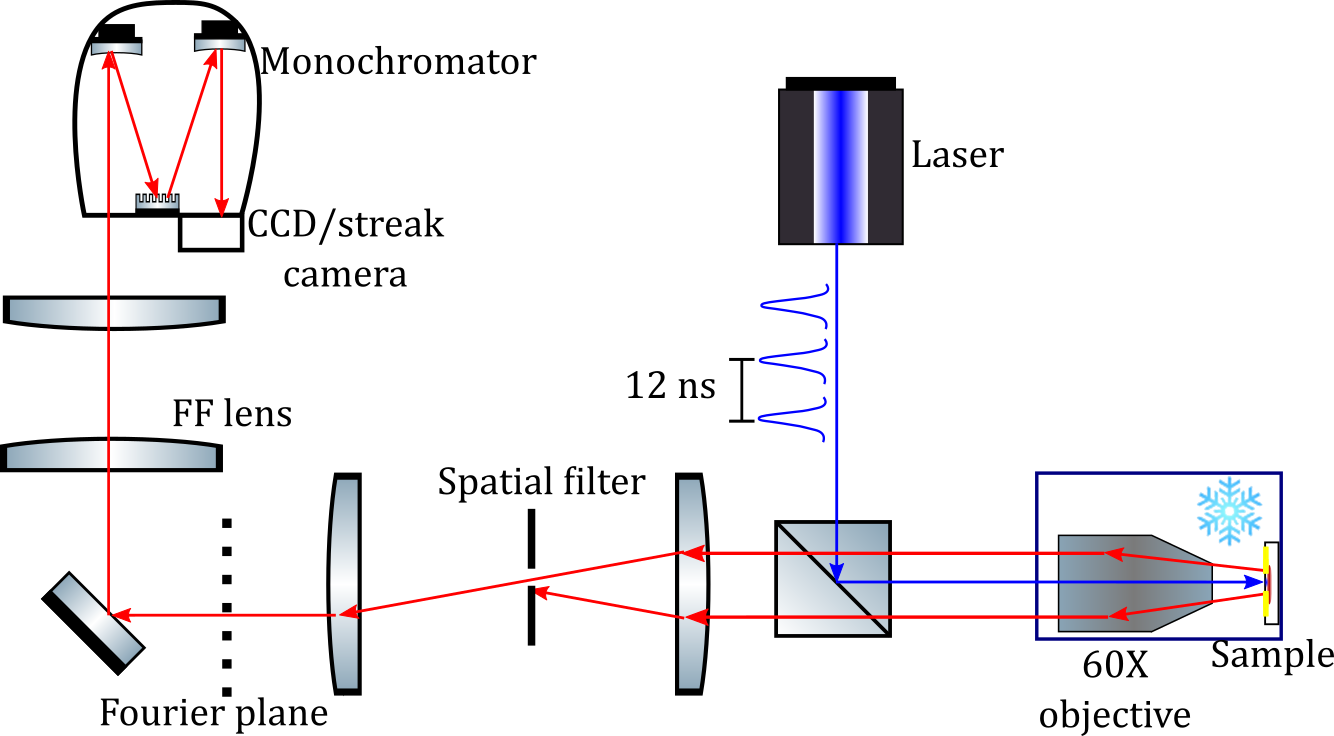}
    \caption{Schematic representation of the experimental setup. The real space image of the sample emission is reconstructed at a given longitudinal position in order to apply a spatial filter, upon using a regulable micrometric slit. Afterwards the image of the real (Fourier) space is reconstructed on the camera by removing (inserting) the Far Field lens (FF lens).}
    \label{setup}
\end{figure}

For all the optical characterization the sample is kept at cryogenic temperature of $4$ K. All the PL measurements are realized in a confocal configuration, using a $100$ fs pulsed laser with a repetition rate of $80$ MHz, tuned at $1.59$ eV ($\approx780$ nm) to excite the sample out of resonance. The detection system allows to reconstruct either real or Fourier spaces in a Charge Coupled Device (CCD) coupled to a monochromator $70$ cm long with a diffractive grating with either $600$ or $1800$ lines per mm. This configuration gives spatial, angular and spectral resolution. An image of the real space is reconstructed before the CCD in order to apply a spatial filter by using a slit, enabling the selection of one or both gratings (as depicted in the panels d and h of Fig.~5 of the main text). A complete scheme of the experimental setup is displayed in Fig.~\ref{setup}. The residual laser signal is suppressed with a long-pass spectral filter at $1.55$ eV ($\approx800$ nm). The time resolved images are performed in the same configuration, but directing the signal into a streak camera after passing the monochromator. The temporal reconstruction of the far field resolved in energy is made by moving vertically the focusing lens with a motorized station before it reaches the streak camera.\\

The energy dispersion is measured by imaging the Fourier plane of the objective lens that collects the light from the sample, as it is usually done when measuring polariton emission. The calibration of the emission wavevector in the propagation direction z depends on the magnification of such plane in the CCD camera and the objective lens working distance. This setup allows to individuate the angle ($\theta$) at which the light is emitted from the grating, which is related to the wavevector by $k_z=2\pi Sin(\theta)/\lambda_o$, where $\lambda_o$ is the vacuum wavelength. Since the light is being extracted from the waveguide by the grating, exploiting the diffraction effect that couples the guided modes to radiative modes, the total wavevector of the confined mode associated to the detected light is given by \cite{Xiao2013}:
$\beta=k_z+2\pi/\Lambda$
Where $\Lambda$ is the grating periodicity.

\section{Calculation of waveguide modes profile and grating coupled mode analysis}
\begin{figure}
    \centering
    \includegraphics[width=0.55\columnwidth]{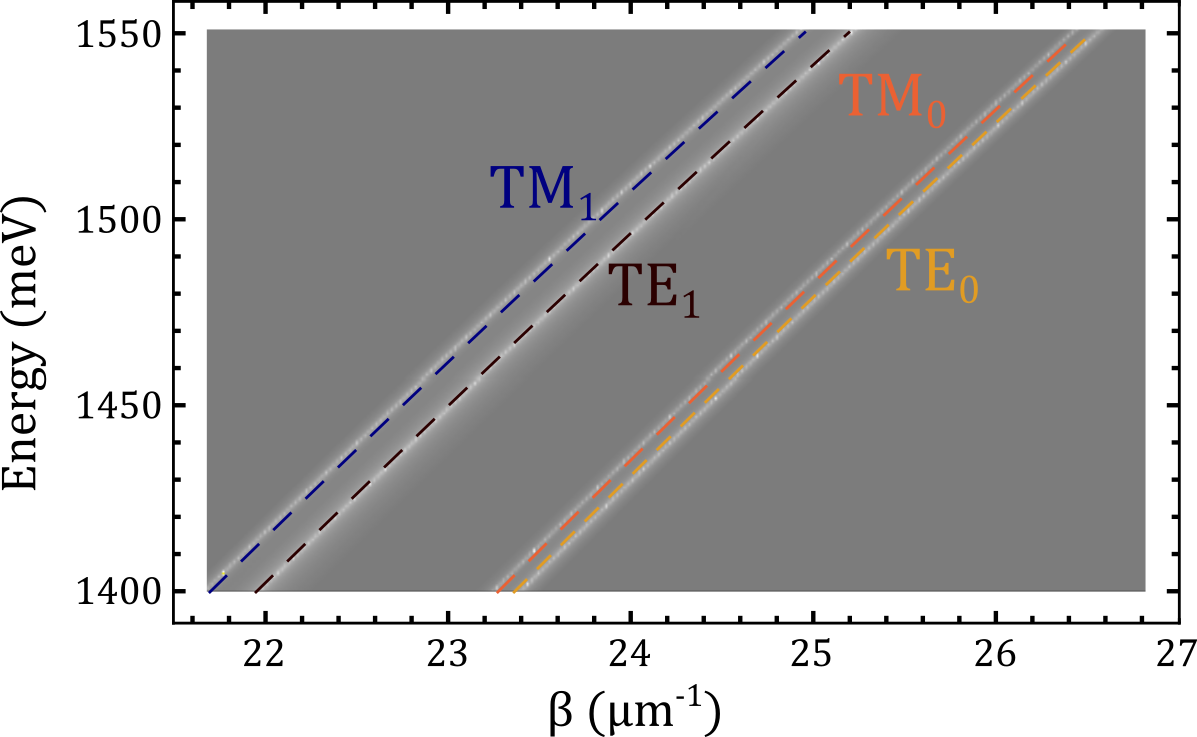}
    \caption{Energy dispersions of the four guided modes supported by the structure, calculated using two different theoretical approaches: the background image (with the modes represented as white lines) corresponds to the dispersion calculated using the transfer matrix method, considering the internal structure of the waveguide, i.e., the 12 pairs of 20 nm layers of Al$_{0.4}$Ga$_{0.6}$As and GaAs. The dashed lines are calculated with the semi-analytical method illustrated in this section, assuming a mean effective index for the entire waveguide thickness. The good agreement between both calculations, validates the approximation of effective refractive index in the theoretical formalism used along the work.}
    \label{TM}
\end{figure}

To perform theoretical calculations on the perturbation induced by the grating on a guided mode, we use a semi-analytical model illustrated in Ref.~\cite{Yariv2007}. It allows to find expressions for the electric and magnetic field distribution for an asymmetric waveguide composed by three homogeneous media. The periodic nature of our structure (12 pairs of QW+barrier) is taken into account by means of an effective index  ($n_{WG}^{eff}$) that weights the contribution of the two constituent materials: $Al_{0.4}Ga_{0.6}As$ ($n=3.30$) and GaAs ($n=3.54$). This approximation is justified by the fact that the layers are 20 nm thick, a size more than one order of magnitude below the wavelength of the photonic mode inside the structure ($\lambda\sim240$ $nm$). The system is then operating in the ``long wavelength" regime, in which the wave does not resolve in the structure \cite{Rytov1956,Brauer1994}. As a further validation of this approach, we compared the energy dispersion of the photonic modes by using a transfer matrix formalism that takes into account the periodicity of the waveguide. The results, presented in Fig.~\ref{TM} confirm the pertinence of the  formalism we use. They show a very good agreement between the two methods. However, the expressions obtained in the analytic formalism have the advantage to permit an accurate description of the interaction of the guided modes with the grating, hence, this is the formalism employed along the work.\\
In our calculations we use the expressions corresponding to an asymmetric waveguide of refractive index $n_2>n_3>n_1$ defined in the spatial range $-t<x<0$, with modes propagating in the $z$ direction and infinitely extending along $y$. For this system the refractive indices correspond to: $n_1=n_{ITO}$, $n_2=n_{WG}^{eff}$ and $n_3=n_{Al_{0.8}Ga_{0.2}As}$. The magnetic field ($\mathcal{H}$) spatial profile of the $TM_0$ mode can be written as:

\begin{equation}\mathcal{H}_y(x)= \begin{cases} 
      -C~\frac{n_1^2 h}{n_2^2 q}exp(-q x) & 0\leq x \\
      C~\left(\frac{n_1^2 h}{n_2^2 q}cos(hx)+sin(hx)\right) & -t\leq x\leq 0 \\
      C~\left(\frac{n_1^2 h}{n_2^2 q}cos(ht)+sin(ht)\right)exp(p(x+t)) & x\leq -t 
   \end{cases}
\end{equation}

Where the constants $q$, $p$ and $h$ are chosen in order to satisfy the boundary conditions of the magnetic and electric fields and the wave equation for the electric field. The constant $C$, instead, determines the field's intensity. The exciton resonance in the quantum well is taken into account by using a model of Lorentz oscillator \cite{Andreani1994,Liscidini2011}, in which the permittivity of the system in strong coupling is given by:
\begin{equation}
    \epsilon(\omega)=\epsilon_{\infty}\left[1+\frac{\omega_{LT}}{\omega_{ex}-\omega-i\gamma}\right]
\end{equation}
Where $\epsilon_{\infty}$ is the background dielectric constant, $\omega_{ex}$ is the resonant energy, $\gamma$ is the damping rate, and $\omega_{LT}$ is the longitudinal/transverse splitting.\\ 

The insertion of a grating of refractive index $n_g$, height $h$, pitch factor $\Lambda$ and filling factor $a$ on top of the slab waveguide, can be treated as a perturbation of the polarization vector that couples a propagating mode with a counter-propagating one, by using the coupled-mode theory \cite{Yariv2007,Yariv1973}. In our case, the perturbation is a periodic binary modulation of the refractive index, so it can be expanded in the Fourier basis as:
\begin{eqnarray}
\Delta n^2(x,z)=\Delta n^2(x) \sum_{l=-\infty}^{\infty}a_l~exp(i 2\pi l z/\Lambda)\\
\Delta n^2 (x)=\begin{cases}(n_g^2-n_1^2) & 0\leq x \leq h\\
0 & elsewhere
\end{cases}\hspace{2cm}a_l=\begin{cases}
\frac{1}{\pi l}sin(\pi l a) & l\neq 0\\
\end{cases}
\end{eqnarray}
If a Fourier component $a_q$ of the grating's spatial function is comparable with a guided mode's wavevector, i.~e., if $\beta_s\approx l\pi/\Lambda$, the grating will couple the propagating mode with a counter-propagating one with opposite wavevector, following the behavior given by the equations system:
\begin{eqnarray}
    \frac{dA_s^-}{dz}=i\frac{\omega \mu}{4}a_q A_s^+e^{i2(\pi q/\Lambda-\beta_s )z} \left(\frac{\epsilon_{Au}}{\epsilon_1}-1\right) \int_{0}^{h}
    \left[\mathcal{H}^s_y(x)\right]^2dx\\
    \frac{dA_s^+}{dz}=-i\frac{\omega \mu}{4}a_q A_s^-e^{-i2(\pi q/\Lambda-\beta_s )z} \left(\frac{\epsilon_{Au}}{\epsilon_1}-1\right) \int_{0}^{h}
    \left[\mathcal{H}^s_y(x)\right]^2dx
\end{eqnarray}

Which can be written in a synthetic way as:

\begin{eqnarray}
    \frac{dA_s^-}{dz}=\kappa A_s^+ e^{-2\Delta\beta z}\\
    \frac{dA_s^+}{dz}=-\kappa A_s^- e^{2\Delta\beta z}
\end{eqnarray}
Where $\lvert A^{+}(z)\lvert^2$ and $\lvert A^{-}(z)\lvert^2$ are the transverse field intensities of the propagating and counter-propagating modes, respectively, at the position $z$,  and the constants $\Delta \beta$, and $\kappa$ are given by:

\begin{equation}
    \Delta\beta=\beta_s-\pi q/\Lambda \hspace{0.7cm}  
\kappa=i\frac{\omega \mu}{4}a_q \left(\frac{\epsilon_{Au}}{\epsilon_1}-1\right) \int_{0}^{h}
    \left[\mathcal{H}^s_y(x)\right]^2dx
\end{equation}

\section{Effect of the gold grating on the TM polaritonic mode}
\begin{figure}
    \centering
    \includegraphics[width=0.75\columnwidth]{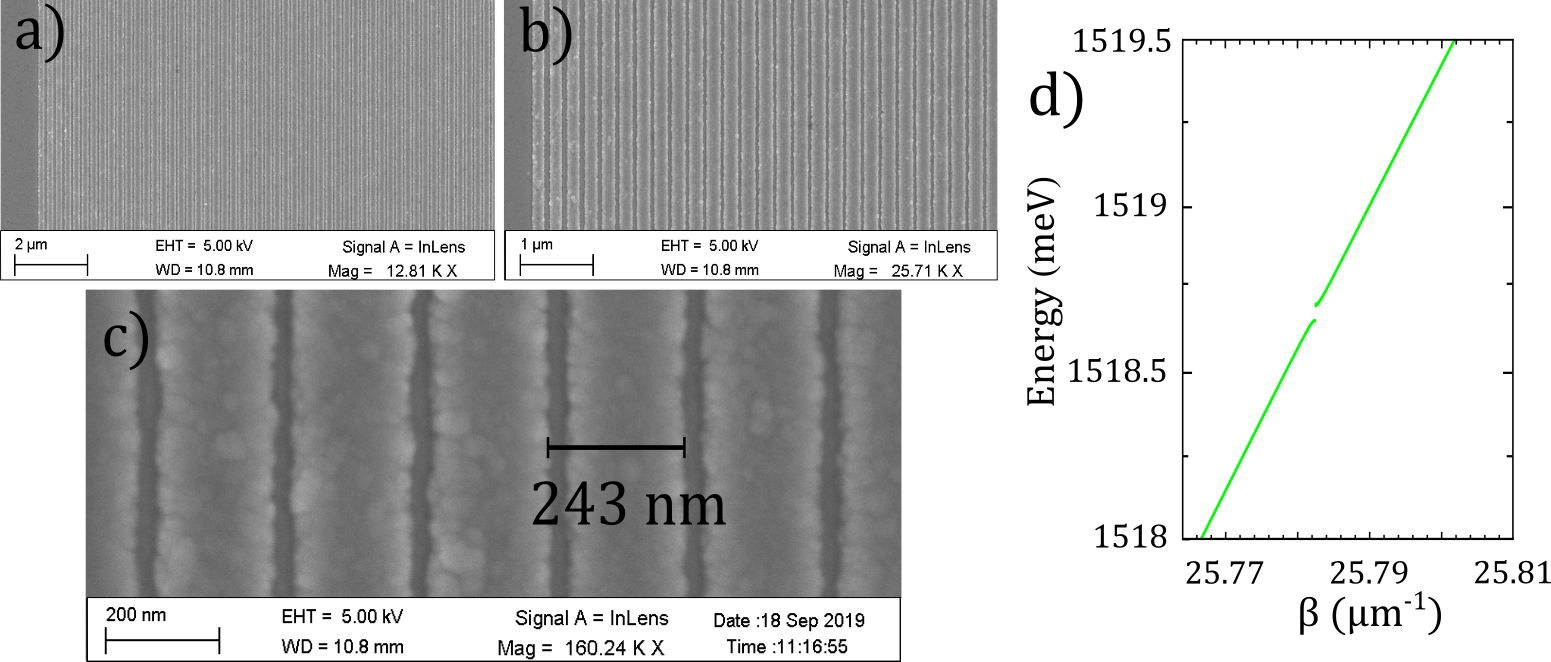}
    \caption{\textbf{a-c} SEM images of a representative gold grating in the structure at three different magnifications: 13k (a), 26k (b) and 160k (c). Panels a and b show a proximity effect in the external part of the grating, due to the dense patterning of the structure. However, as observed in the same panels, after a few periods, the wire width regularizes. The more magnified image (panel c) give details on the structure parameters. It is characterized by a pitch of $243$ nm and a filling factor comprised between $~0.72$ and $~0.85$. \textbf{d} A theoretical analysis indicates that the dispersion under the grating region is perturbed, and, as a consequence, an energy gap appears in the guided mode, which implies that the grating can act as a semi-reflective mirror for incoming WG modes with those energies.}
    \label{photo}
\end{figure}

\begin{figure}
    \centering
    \includegraphics[width=0.75\columnwidth]{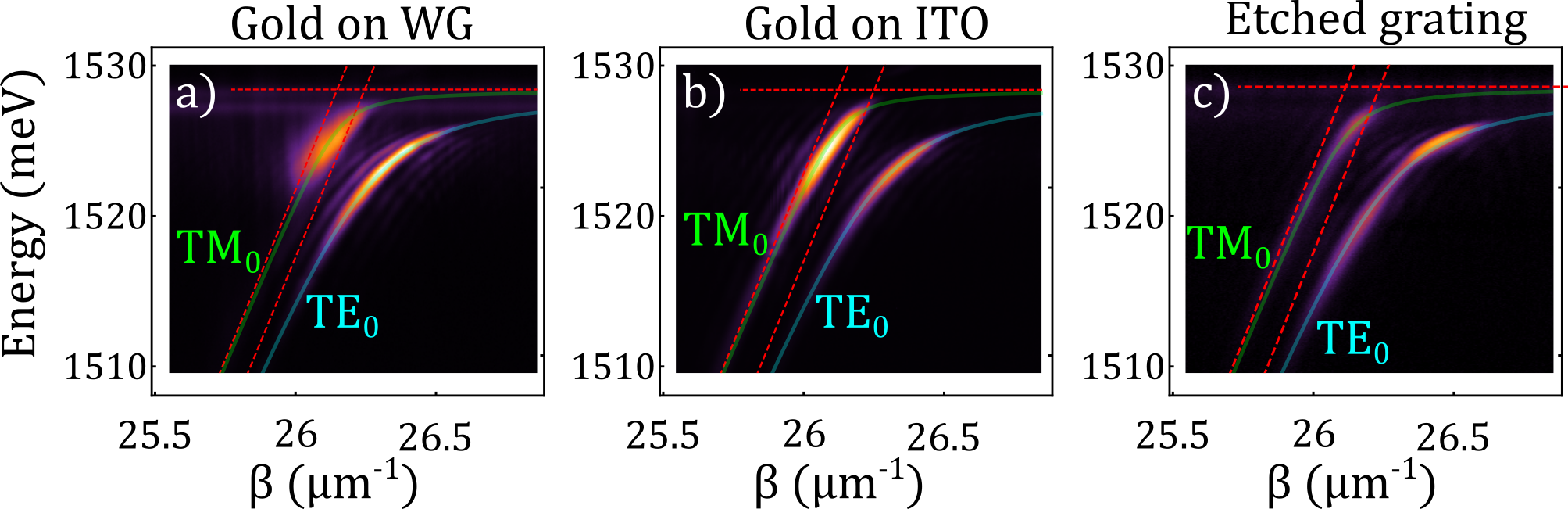}
    \caption{Comparison of the PL spectrum of the guided modes extracted through three different gratings: \textbf{a} A metallic grating on top of the WG slab. Possible plasmonic effects manifest in the TM dispersion as a larger linewidth on the measured PL extracted from this grating. \textbf{b} A less perturbative grating on top of the ITO layer. The effect of the spread in $\beta$ is suppressed. \textbf{c} To verify that no plasmonic effect is altering the detected WG dispersion, we perform the extraction from a dielectric grating etched on top of the slab surface.}
    \label{modes_ito}
\end{figure}

As a consequence of the periodic modulation of the refractive index, an energy gap centered at $\omega(\beta_s=\pi q/\Lambda)$ is generated. As illustrated in the main text, we use gold gratings $30$ nm high with a filling factor comprised between $0.72$ and $0.85$ and a pitch of $243$ nm. Figures \ref{photo}a-c show SEM images of a representative grating, taken at three different magnifications. In panels a and b it is possible to appreciate the irregular wire widths of the first few lines of the grating. This is a consequence of the higher dose at which the more dense features are exposed, a well known particularity of the nano-lithography technique used to fabricate the gratings (electron beam lithography), known as ``proximity effect" \cite{Kim2013}. The region affected comprises less than $1$ $\mu m$, though; a negligible portion compared to the full grating length ($100$ $\mu m$). An image taken with greater magnification in the central region of the grating is displayed in Fig.~\ref{photo}c, where the regular pitch can be appreciated. The small imperfections in the edges of each wire are in the order of units of nanometers, a scale well below the effective wavelength of the polaritonic mode ($\sim240$ nm) by around two orders of magnitude, making the irregularities negligible for the mode analysis. Panel d of the same figure displays the dispersion relation calculated according to the method illustrated above. The existence of the energy gap implies that the grating acts as a semi-reflective mirror in that spectral range.
Figure 1 of the main text and Fig.~\ref{modes_ito}a show that the TM mode is broader than the TE mode, i.e., it is extracted with a lower angular resolution. To verify that this spread is not an intrinsic property of the waveguide, but rather a consequence of the interaction of the TM mode with the gold grating, the PL experiment is replicated, but this time, extracting the signal from a grating on top of the ITO, i.e., a less perturbative grating. As evidenced in Fig.~\ref{modes_ito} b, in this case both modes have comparable linewidths. This indicates that the broadening of the TM mode is a consequence of its interaction with the grating fabricated on top of the slab. It also indicates that the grating is much more perturbative for the TM mode than for the TE, in spite of the fact that the interaction is exclusively through the vanishing tail. As discussed in the main text, the stronger effect on the TM mode with respect to the TE, can be due to several effects such as a plasmonic resonance of the nanowires composing the grating or the proximity of the band gap to the excitonic resonance, the TM gap being closer to the energy range where the PL is more intense. The PL spectra extracted from gold gratings in contact with the slab and on the ITO capping layer are compared with a third PL measurement taken from a fully semiconductor grating. The results displayed in Fig.~\ref{modes_ito}c confirm that the system is working in the strong coupling regime with the excitons of the QWs inside the WG, and discard the possibility of an apparent strong coupling induced by a coupling of the guided modes with a plasmonic resonance of the metallic grating.
A direct measurement of the energy gap introduced by the gratings is done by extracting the PL through a grating $400$ $\mu m$ long, covering its first $100$ $\mu m$ with a mask in the near-field as it is described in the main text.

\section{Properties of the laser emission}
\begin{figure}
    \centering
    \includegraphics[width=0.95\columnwidth]{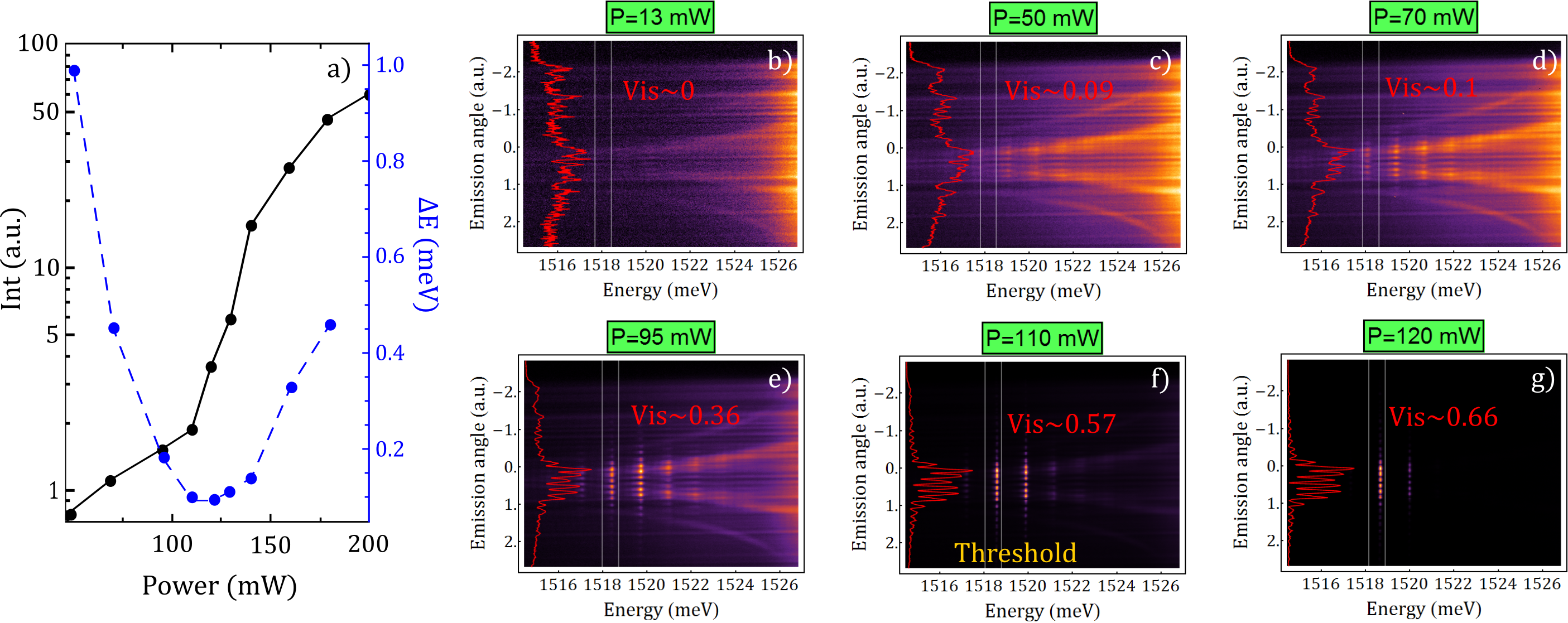}
    \caption{\textbf{a} Emitted intensity (black solid line) and FWHM (blue dashed line) of the lasing mode upon increasing the pump power. A reduction in the linewidth at the lasing threshold is clearly visible. \textbf{b-g} Overlap in k space of the laser emission coming from the two gratings separated $100$ $\mu m$, and for different pumping powers. The interference pattern results from the overlap of the coherent beams emitted by the two gratings. The white lines indicate the region corresponding to the profile depicted in red. The visibility of the interference fringes is deduced and specified for each pump power.}
    \label{coherence}
\end{figure}

Figure \ref{coherence} shows the main properties of the polariton laser demonstrated in this work. The panel \ref{coherence}a shows the total emitted intensity (black solid line) and full width at half maximum (FWHM, blue dashed line) of the lasing mode as a function of the pump power (see also Figures 3 and 4 of the main text). The FWHM is reduced when the pump power is increased and it reaches a minimum just above the lasing threshold. This behaviour is well known in systems showing polariton lasing and it is explained as an increase of the coherence time above the lifetime of the polariton mode. At higher powers, a small increase of the FWHM is observed, which can be due to polariton-polariton and polariton-reservoir interactions. The panels b to g show the overlap in the k space of the light emitted by the two gold gratings. Below threshold, the pump generates population spreading in all directions (of the waveguide plane) and in both modes, TE and TM (13 mW and 50 mW), but no interference can be observed when overlapping the emission from both gratings in the Fourier space. As the power increases, the modes of the FP cavity acquire enough population to become visible (70 mW and 95 mW), but the degree of coherence is still very low. The system then reaches the threshold power (110 mW) until the lasing mode is massively populated, achieving high coherence (120 mW). The figure then illustrates how the system passes from an incoherent state (as expected for a polaritonic system pumped out of resonance in the low power regime), to a gradual enhancement of the coherence as the pump power increases.

\section{Power threshold for different cavity lengths}
\begin{figure}
    \centering
    \includegraphics[width=0.99\columnwidth]{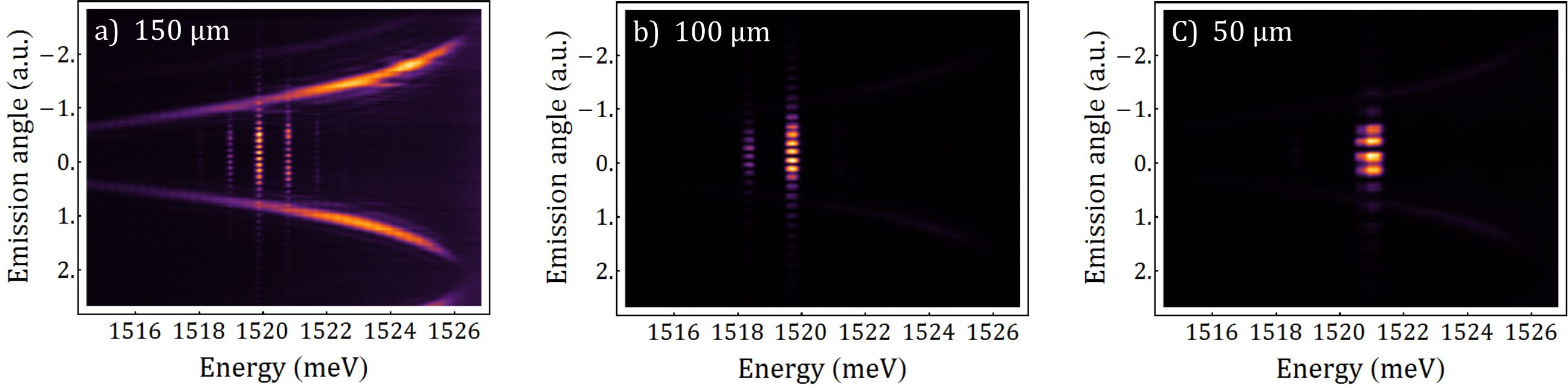}
    \caption{Emission collected from pairs of gratings separated by 150 $\mu m$ (a), 100 $\mu m$ (b) and 50 $\mu m$ (c). In the three cases the pump power is 150 mW, but in each case the system is in a different regime, demonstrating a dependence of the threshold power on the cavity length. The integrated emission intensity (panel d) confirms the different state of each system and in particular that the system in panel a is at the lasing threshold.}
    \label{power_distance}
\end{figure}

Regarding the power dependence on the gratings spacing, we compared the emission from FP cavities of three different lengths (50 $\mu m$, 100 $\mu m$ and 150 $\mu m$), pumping in each case close to the threshold power (150 mW) with the same spot size. The results, displayed in  Fig.~\ref{power_distance}, show that for the same pump power the system can be found in different regimes, depending on the spacing between gratings. For a distance of 150 $\mu m$ the system is near the lasing threshold, for 100 $\mu m$ the system is above the lasing threshold, and for the distance of 50 $\mu m$ the system is above the lasing threshold and in the interactive regime, where the interactions start to affect the laser linewidth (the characterization of the three regimes for a cavity of 100 $\mu m$ are presented in Fig.~\ref{coherence}). The trend of the integrated emission intensity (panel d) is in agreement with this interpretation. This behavior is expected, since the cavity length changes the total dissipation of the FP mode: the shorter the cavity the lower the losses. The pump area is the same in the three cases, though. The consequence is a modification of the ratio gain/losses, and therefore, of the threshold power.

\section{Appliance of an electric field}
\begin{figure}
    \centering
    \includegraphics[width=0.35\columnwidth]{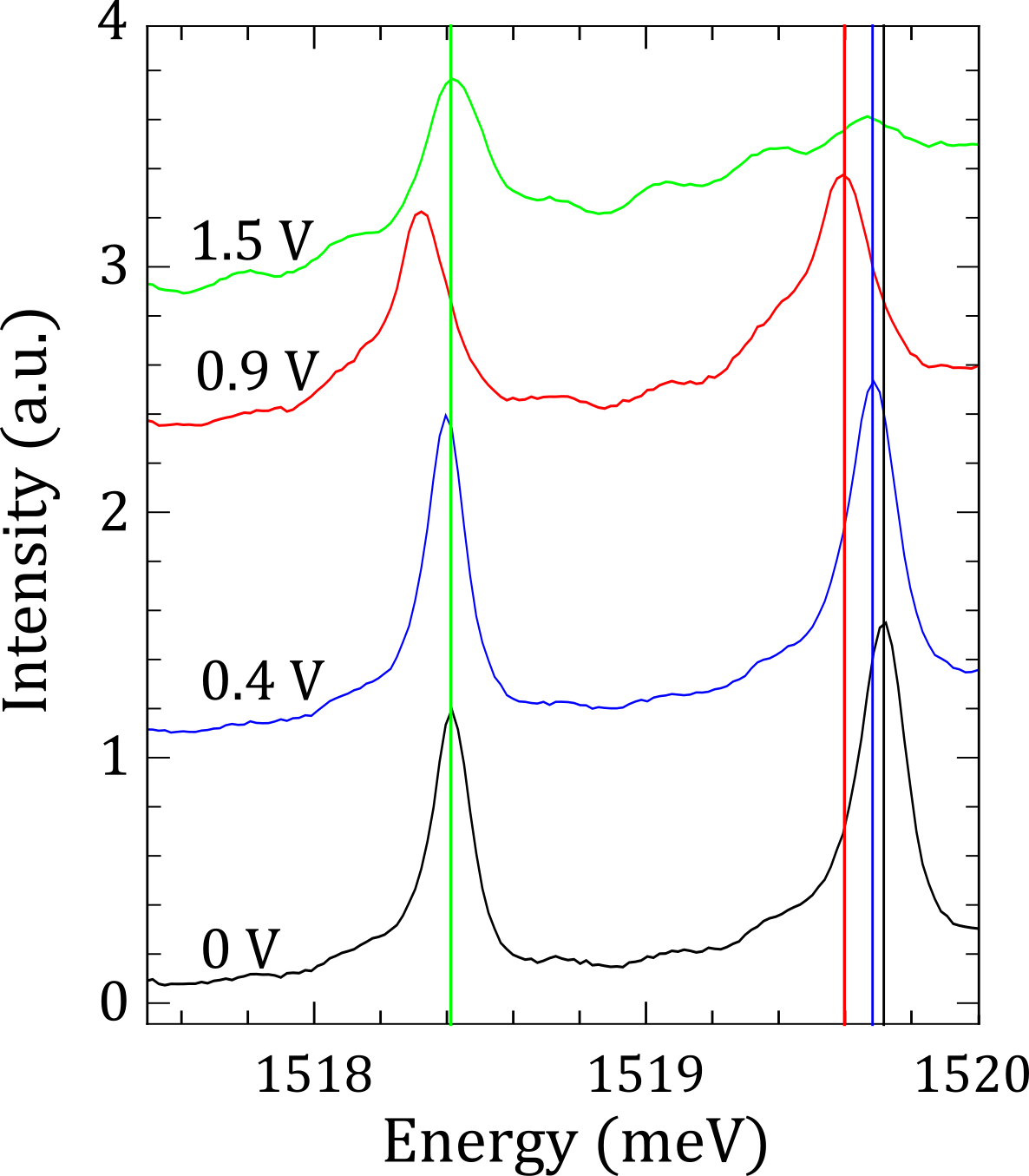}
    \caption{By zooming the region of the cavity modes in Fig. 4a of the main text, it is possible to identify the variations in the energy of the lasing mode (indicated by the vertical lines) for the different applied voltages. In the cases of $0.4$ V and $0.9$ V, the variation is small ($0.02$ meV and $0.12$ meV, respectively), while in the case of $1.5$ V there is an overlapping of the exciton with the bluest cavity mode; in this case the polariton lasing effect takes place in the consecutive one.}
    \label{modes}
\end{figure}

The effect of the electric field is the energy shift of the exciton peak by means of the Stark effect. As illustrated in Fig. 4a in the main text, the electric field shifts the excitonic resonance towards the modes of the Fabry-Perot resonator. As the exciton gets closer, the absorption, and hence, the quality factor of each mode is affected. At a voltage of $1.5$ V the exciton peak is overlapped with the Fabry-Perot mode, increasing noticeably the absorption at this energy, and hence, reducing drastically the quality factor of the mode. Fig.~\ref{modes} shows both effects. On the one hand, for applied voltages of $0.4$ V and $0.9$ V there is a shift of $0.02$ meV and $0.12$ meV, respectively. This demonstrates the possibility of a very fine tuning of the laser energy with the applied field. The spectrum corresponding to the highest applied field ($1.5$ V) shows how the change in the absorption peak affects the quality factor of the confined modes, inducing a spectral jump in the lasing energy, and hence, opening the possibility to use the applied electric field also as a switch to choose the desired lasing mode.

\end{document}